\documentclass[aps,prapplied,reprint,twocolumn,groupedaddress]{revtex4-2}
\usepackage[colorlinks,linkcolor=blue,urlcolor=blue,anchorcolor=blue,citecolor=blue]{hyperref}
\usepackage{amsmath}
\usepackage{graphicx}
\usepackage{dcolumn}
\usepackage{mathrsfs}
\usepackage{amssymb}
\usepackage{amsfonts,multirow}
\usepackage{bm,changes}
\usepackage{color}
\usepackage{ulem}
\usepackage{changes}

\begin{document}
\title{Dissipation-Coupled Bosonic Quantum Battery Protected by a Dark State}
\author{Hao-Long Zhang$^{1}$}
\address{$^{1}$Fujian Key Laboratory of Quantum Information and Quantum\\
Optics, College of Physics and Information Engineering, Fuzhou University,\\
Fuzhou 350108, China\\}
\author{Wen Ning$^{1}$}
\email{ningw@fzu.edu.cn}
\address{$^{1}$Fujian Key Laboratory of Quantum Information and Quantum\\
Optics, College of Physics and Information Engineering, Fuzhou University,\\
Fuzhou 350108, China\\}
\author{Zhen-Biao Yang$^{1}$}
\email{zbyang@fzu.edu.cn}
\address{$^{1}$Fujian Key Laboratory of Quantum Information and Quantum\\
Optics, College of Physics and Information Engineering, Fuzhou University,\\
Fuzhou 350108, China\\}
\author{Shi-Biao Zheng$^{1}$}
\email{t96034@fzu.edu.cn}
\address{$^{1}$Fujian Key Laboratory of Quantum Information and Quantum\\
Optics, College of Physics and Information Engineering, Fuzhou University,\\
Fuzhou 350108, China\\}

\begin{abstract}
Quantum batteries can in principle store and deliver energy beyond classical limits, but environmental dissipation during charging erodes both stored energy and extractable work. Dark-state protection, which confines the battery to a subspace decoupled from the bath, suppresses this degradation without external driving or feedback, yet so far it has been demonstrated only in discrete-variable systems with qubits or qutrits. Here we show that dark-state protection is achievable with two bosonic modes only: a battery and a charger coupled solely through a common dissipative reservoir, whose one canonical combination decays while the other remains dark and decoupled. The stored energy grows with the charger's initial photon number, and when the charger carries coherence, essentially the entire stored energy becomes extractable as work. Replacing the linear channel by a two-photon channel acting on the decaying mode alone preserves this protection while shortening the charging time as the photon number grows, so that the peak power becomes superextensive and approaches $\bar n^{2}$. Our results extend dark-state battery protection from many-body spin platforms to a minimal, few-body bosonic setting. 
\end{abstract}

\maketitle

\vskip0.5cm

\narrowtext

\section{Introduction} 
Coherence and entanglement are established resources for quantum technologies
\cite{Nielsen.2000,PhilosTransRSocA.361.1655,JPhysCondensMatter.18.V1}, and quantum
thermodynamics \cite{Entropy.15.2100,NatPhys.11.118,ContempPhys.57.545,JPhysA.49.143001,NewJPhys.19.010201,RepProgPhys.82.114001,Binder.2019,ContempPhys.60.164} has motivated quantum batteries
(QBs) as a route to optimize energy storage and work extraction \cite{PhysRevE.87.042123,PhysRevLett.111.240401,NatCommun.5.4185,NewJPhys.17.075015,PhysRevA.97.022106,NewJPhys.22.063057,PhysRevLett.122.210601,PhysRevA.101.032115,RevModPhys.96.031001}. By exploiting coherence and other nonclassical features, QBs aim at operating principles fundamentally different from those of conventional batteries.

Numerous protocols \cite{PhysRevLett.134.130401,PhysRevA.105.062203,PhysRevLett.131.240401,PhysRevLett.132.090401,PhysRevLett.122.210601,PhysRevLett.120.117702,PhysRevLett.136.060401,PhysRevA.113.032220,PhysRevE.111.064117,PhysRevLett.136.180401} and experiments \cite{SciAdv.8.eabk3160,LightSciAppl.15.168,PhysRevA.106.042601,QuanSciTech.7.025020,PhysRevLett.131.260401,QuanSciTech.7.045018,PhysRevAppl.24.054033} have been proposed to achieve
performance beyond classical limits, i.e., a quantum advantage
\cite{PhysRevLett.118.150601}. Many-body QBs, with the Dicke battery as the paradigmatic example
\cite{PhysRevLett.120.117702,PhysRevB.109.235432,PhysRevB.102.245407,PhysRevA.110.042419,Batteries.9.197},
were proposed as candidates for superextensive charging
\cite{PhysRevA.97.022106,PhysRevLett.133.197001,PhysRevB.100.115142,PhysRevA.101.032115,PhysRevE.105.054115,PRXQuantum.5.030319,PhysRevLett.134.180401}. This enhancement, however, originates from the collective interaction
\cite{FrontPhys.10.1097564} rather than from a genuinely quantum feature such as entanglement, and it is therefore
insufficient to demonstrate a quantum advantage. Models based on the Sachdev-Ye-Kitaev (SYK) Hamiltonian have been proposed to exhibit such an advantage \cite{PhysRevLett.125.236402,JHighEnergPhys.2020.67},
but they are difficult to implement experimentally. Bosonic QBs \cite{CommunPhys.6.322,PhysRevA.108.052213,PhysRevA.110.022226,PhysRevA.112.032209,PhysRevB.98.205423,PhysRevB.108.L180301,PhysRevE.112.044143,PhysRevLett.134.240403,PhysRevLett.134.220402,Quantum.2.61,SciChinaPhysMechAstron.69.260312,ApplPhysLett.123.034001,PhysRevLett.137.180404,arXiv.2511.20569,arXiv.2611.06712, arXiv.2610.35734} have recently attracted attention in
this context: thanks to their infinite-dimensional Hilbert space, they can reach
superextensive charging already in few-body systems \cite{PhysRevLett.134.240403}.

On the other hand, quantum systems inevitably couple to their environment, which causes
decoherence and degrades the performance of quantum technologies. Protocols have been proposed to stabilize or even enhance the performance of an open QB, including non-Markovian environment engineering \cite{NewJPhys.22.083007,PhysRevA.104.032207,PhysRevA.102.052223,PhysRevE.104.064143}, quantum Zeno
stabilization \cite{PhysRevResearch.2.013095}, and dark-state protection
\cite{arXiv.2605.17700,PhysRevApplied.14.024092,PhysRevA.113.012617,arXiv.2611.15416,PhysRevA.114.022420,PhysRevA.114.022455,arXiv.2610.36704,arXiv.2608.09693,arXiv.2606.27403}. Among them, dark-state-protected QBs have been realized with discrete-variable systems, where spins are restricted to qubits or qutrits. However, dark-state-protected QBs remain unexplored for continuous-variable (bosonic) systems.

We study a bosonic QB in which the battery and charger are two modes coupled solely
through a common dissipative reservoir. This configuration requires neither external driving nor
feedback: the symmetric combination of the two modes decays, whereas the antisymmetric combination is dark
and decoupled from the reservoir, so that the battery is passively protected. For a charger prepared
in a Fock state, we obtain the exact steady state of the battery, a binomial mixture of photon-number
states, and show that its stored energy grows linearly with the initial photon number, whereas its
ergotropy is carried entirely by the population inversion and vanishes for small photon numbers.
Preparing the charger in a coherent state instead leaves a finite coherence in the steady state and
makes almost the entire stored energy extractable. We further evaluate the charging power, whose
peak scales linearly with the photon number. To achieve the superextensive charging, we consider a two-photon channel and perform numerical simulation; the charging power then scales superextensively with the photon number of the charger. Our simulations show that the charging speed
is governed by the dissipative dynamics rather than by the statistics of the charger. In contrast to spin-based QBs, which require many constituents, our scheme increases the stored energy by increasing the photon number of a single charger mode, and both single-photon and two-photon channels are within reach of circuit QED.


\section{Dissipative dynamics of the bosonic quantum battery}
\begin{figure}[t]
\includegraphics{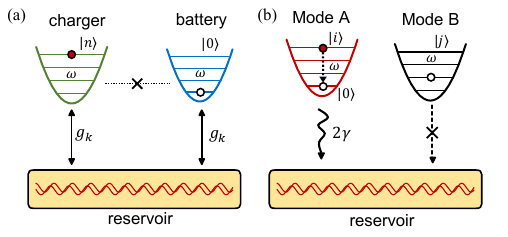}
\caption{(a) Charging setup and canonical modes. The battery mode $b$ is prepared in the vacuum and the
charger mode $c$ is prepared in a Fock state with $n$ photons. The two modes are not directly coupled; but both interact
with a common reservoir through the collective coupling $g_k$. (b) In the canonical-mode basis $A=(b+c)/\sqrt{2}$ and $B=(b-c)/\sqrt{2}$, only the bright mode $A$ couples to
the reservoir, at rate $2\gamma$, whereas the dark mode $B$ is decoupled.}
\label{fig:model}
\end{figure}

\vspace*{-6pt}
We consider a battery mode $b$ prepared in the vacuum state and a charger mode $c$ prepared with photon number $n$ [Fig.~\ref{fig:model}(a)], they interact only through a common
dissipative reservoir. In a frame rotating at two oscillators' resonant frequency $\omega$, the interaction Hamiltonian is ($\hbar$ = 1)
\begin{equation}
H_I(t) =\sum_{\mathbf k} g_{\mathbf k}\big(c^\dagger+b^\dagger\big)a_{\mathbf k}\,
e^{-i(\nu_{\mathbf k}-\omega)t}+\mathrm{h.c.},
\end{equation}
where $b$ ($b^\dagger$) and $c$ ($c^\dagger$) are the annihilation (creation) operators of the
battery and the charger, respectively, $a_{\mathbf k}$ ($a_{\mathbf k}^\dagger$) annihilates (creates) a reservoir
excitation with wave vector $\mathbf k$ and frequency $\nu_{\mathbf k}$, and $g_{\mathbf k}$ is its
coupling strength to the two oscillators. 

Assuming initially uncorrelated system and reservoir, and applying the
Born--Markov approximation, the charging dynamics obeys the Lindblad master equation
\begin{equation}\label{master}
      \begin{split}
            \dot\rho(t)=&-i\omega\big[b^{\dagger}b + c^{\dagger}c, \rho(t)\big] \\&+ \gamma
\big\{(\bar n+1)\mathcal{D}[b + c]\rho(t) + \bar n\,\mathcal{D}[b^{\dagger} + c^{\dagger}]\rho(t)
\big\},
      \end{split}
\end{equation}
where $\mathcal{D}[O]\rho\equiv O\rho O^\dagger-\{O^\dagger O, \rho\}/2$, $\gamma$ is the system--reservoir coupling rate, and
$\bar n=1/[\exp(\hbar\omega/k_{B}T)-1]$ is the mean thermal occupation of the reservoir. We work
in the zero-temperature limit, $\bar n\to0$, where the second term of Eq.~(\ref{master}) vanishes.

The dark states are conventionally identified by introducing the canonical modes [Fig.~\ref{fig:model}(b)]
\begin{equation}\label{canonical}
            A = (b + c)/\sqrt{2},\
            B = (b - c)/\sqrt{2}.
\end{equation}
Substituting Eq.~(\ref{canonical}) into Eq.~(\ref{master}), the time evolution of the system becomes
\begin{equation}
      \begin{split}
            \dot\rho(t)=&-i\omega\big[B^{\dagger}B + A^{\dagger}A, \rho(t)\big] \\&+  2\gamma
\big\{(\bar n+1)\mathcal{D}[A]\rho(t) + \bar n\,\mathcal{D}[A^{\dagger}]\rho(t) \big\}.
      \end{split}
\end{equation}
The joint eigenstates of $A^\dagger A$ and $B^\dagger B$ form a complete basis,
\begin{equation}\label{ij}
\vert i,j \rangle_{A,B}=\frac{(A^\dagger)^i}{\sqrt{i!}}\frac{(B^\dagger)^j}{\sqrt{j!}}\vert vac \rangle,
\end{equation}
where $\vert vac\rangle$ is the common vacuum of the two modes, and $i$ and $j$ are the excitation
numbers of the canonical modes $A$ and $B$. The dark states are those with $i=0$. Because $B$ is decoupled from the reservoir, $j$ is conserved during the
evolution; at long times, every component with a given $j$ decays into the
corresponding dark state $\vert 0,j \rangle_{A,B}$. For an initial state $\vert
\psi(0)\rangle=\sum_{i,j}u_{i,j}(0)\vert i,j\rangle$, the steady state is therefore
\begin{equation}
      \begin{split}
        \rho_{ss} =& \sum_j\sum_i \vert u_{i,j} \vert^2 \vert 0,j\rangle_{AB} \langle 0,j\vert \\& +\sum_{i\neq j}u_{0,i}u_{0,j}^*\vert 0,i\rangle_{AB}\langle 0,j\vert,        
      \end{split}
\end{equation}
with $u_{i,j} = u_{i,j}(0)$. Tracing out the charger gives the steady-state battery density matrix
\begin{equation}
      \begin{split}
            \rho_{ss}^{B} = &\sum_j\sum_i \vert u_{i,j} \vert^2 \mathrm{Tr}_{c}\left[\vert
0,j\rangle_{AB} \langle 0,j\vert\right] \\&+ \sum_{i\neq
j}u_{0,i}u_{0,j}^{*}\mathrm{Tr}_{c}\left[\vert 0,i\rangle_{AB}\langle 0,j\vert\right].
      \end{split}
\end{equation}
The first term gives the dark-state populations; the second encodes their
coherences. This coherence is the key resource for ergotropy analyzed in Sec.~\ref{III}. When the initial state has support on at most one dark state, no coherence
survives, and $\rho_{ss}^B$ is a classical mixture of photon-number states.

\section{Ergotropy and charging power}\label{III}

The battery performance is characterized by two figures of merit: the ergotropy
and the charging power. For unitary operations $U$ acting on the
battery alone, the extractable work of a battery state $\rho_B$ with Hamiltonian $H_B$ is bounded
by its ergotropy,
\begin{equation}\label{ergo}
W_B(\rho_B)=E_B(\rho_B)-\min_{U}E_B(U\rho_BU^{\dagger}),
\end{equation}
where $E_B(\rho)=\mathrm{Tr}(H_B\rho)$ and $H_B=\hbar\omega\,b^\dagger b$. The optimal unitary maps $\rho_B$
onto its passive state $\sigma_B$, i.e.\ lowest-energy state that is reachable by unitary
evolution. The passive state is diagonal in the energy eigenbasis with
populations decreasing monotonically with energy,
\begin{equation}
\sigma_B=\sum_{k}r_k|k\rangle\langle k|,
\end{equation}
where the eigenvalues $r_k$ of $\rho_B$ are ordered decreasingly, $r_1\ge r_2\ge\cdots$, and the eigenstates $|k\rangle$ of $H_B$ are ordered by increasing energy.

Two contributions to the ergotropy can be distinguished \cite{arXiv.2605.17700}. The first originates from coherence between energy eigenstates and is quantified by the $\ell_1$ norm of the off-diagonal elements
\begin{equation}
\mathcal{C}_B=\sum_{i\neq j}\left|\langle i|\rho_B|j\rangle\right|.
\end{equation}
The second is the incoherent population inversion, carried by the diagonal part alone,
\begin{equation}
W_B^{P}=W_B(\rho_B^{\mathrm{diag}}),
\end{equation}
where $\rho_B^{\mathrm{diag}}=\sum_j\langle j|\rho_B|j\rangle|j\rangle\langle j|$. The remainder,
\begin{equation}
W_B^{C}=W_B(\rho_B)-W_B^{P},
\end{equation}
is attributed to coherence, so that $W_B=W_B^{P}+W_B^{C}$ by construction.

The charging power is the rate at which the battery energy grows,
\begin{equation}
P(t)=\frac{dE_B}{dt},
\end{equation}
and a protocol is characterized by its peak value
\begin{equation}
P_{\max}=\max_{t}P(t).
\end{equation}
Both quantities follow from the battery density matrix $\rho_B(t)$, obtained by solving
Eq.~(\ref{master}). Below we evaluate them numerically for different initial states of the
charger and use the results to characterize the battery performance.

\begin{figure}[t]
\includegraphics{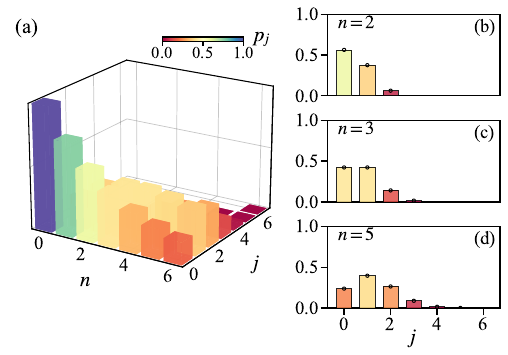}
\caption{Steady-state photon-number distribution $p_j$ of the battery for a Fock-state charger. (a)~$p_j$ versus the charger's initial photon number $n$ and the battery photon number $j$, the color scale is indicated by the color bar. (b)--(d)~$p_j$ for
$n=2$, $3$, and $5$. In (b)-(d), bars and open circles show the numerical solution of Eq.~(\ref{master}) and  the analytic result of Eq.~(\ref{fockdist}), respectively.}
\label{fig:fock}
\end{figure}
\section{Charging performance for different charger states}
\subsection{Charging with the charger in a Fock state}

\begin{figure*}[t]
\includegraphics[width=\textwidth]{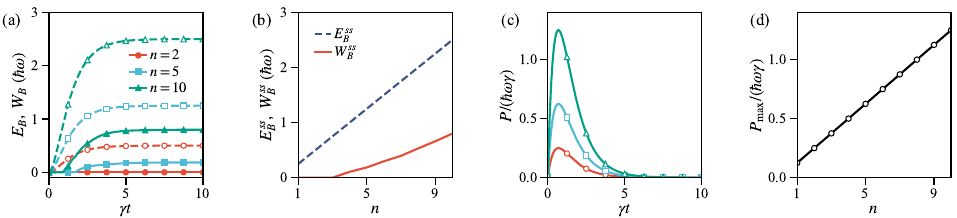}
\caption{Charging performance for a Fock-state charger with $n$ photons.
(a)~Stored energy $E_B(t)$ (dashed) and ergotropy $W_B(t)$ (solid) versus rescaled time
$\gamma t$ for $n=2$, $5$, and $10$; markers distinguish the three $n$ in grayscale. (b)~Steady-state values $E_B^{ss}$ (dashed) and
$W_B^{ss}$ (solid) versus $n$. (c)~Charging power $P(t)=dE_B/dt$ for the same three $n$, with the
colors and markers of (a). (d)~Peak charging power $P_{\max}$ versus $n$. Energies are in units of $\hbar\omega$ and powers in units of
$\hbar\omega\gamma$.}
\label{fig:charge}
\end{figure*}
We first take the charger to be in a Fock state with $n$ photons. In the canonical-mode basis the initial state is
\begin{equation}
\vert\psi(0)\rangle=\vert n\rangle_c\vert 0\rangle_b=\sum_{k=0}^{n}C_k\vert n-k,k\rangle_{A,B},
\end{equation}
so that each component contains $i=n-k$ excitations in the decaying mode $A$ and $j=k$ in the dark mode
$B$. Because only $A$ decays, $\vert n-k,k\rangle_{A,B}$ evolves into the dark state
$\vert 0,k\rangle_{A,B}$, and the steady state is
\begin{equation}
\rho_{ss}=\sum_{k=0}^{n}\vert C_k\vert^{2}\vert 0,k\rangle_{A,B}\langle 0,k\vert .
\end{equation}
Tracing out the charger using Eq.~(\ref{ij}) gives the steady-state density matrix
(see Appendix~\ref{app:steady} for details),
\begin{equation}\label{fockdist}
\rho^{B}_{ss}=\sum_{j=0}^{n}p_j\vert j\rangle\langle j\vert
=\frac{1}{4^{n}}\sum_{j=0}^{n}\binom{n}{j}3^{\,n-j}\vert j\rangle\langle j\vert ,
\end{equation}
where $p_j$ is the photon-number distribution of the battery. The stored energy is
$E_B^{ss}=\hbar\omega\sum_j j\,p_j=n\hbar\omega/4$, and therefore grows linearly with $n$. In many-body QBs, such as spin ensembles, the stored energy and thus the experimental resource overhead scales with the number of spins. By contrast, our scheme requires only an increase in the initial photon
number of the charger. 

Besides the stored energy, the performance of the battery is quantified by the ergotropy. Since $\rho_{ss}^B$ is diagonal in the Fock basis, its coherence contribution vanishes, $W_B^{C}=0$, and the ergotropy is determined solely by the population inversion, $W_B^{P}$. As shown in Fig.~\ref{fig:fock}(a), for $n\le3$, no population inversion occurs [Fig.~\ref{fig:fock}(b)(c)], so the ergotropy vanishes. For $n\ge4$
[Fig.~\ref{fig:fock}(d)], population inversion appears, and the ergotropy becomes finite and increases
monotonically with $n$. Thus the ergotropy can also be raised simply by increasing the initial
photon number of the charger.

We further numerically evaluate the stored energy and the ergotropy as functions of evolution
time and $n$. Both quantities grow monotonically and converge to their steady-state values [Fig.~\ref{fig:charge}(a)], which manifests the
dark-state protection of the battery. The converged stored energy increases linearly with $n$, consistent with $E_B^{ss}=n\hbar\omega/4$. In contrast, the ergotropy vanishes for small $n$; as $n$
increases, it builds up and converges to a steady-state value, that itself grows with $n$, from
$0.18\hbar\omega$ at $n=5$ to $0.80\hbar\omega$ at $n=10$ [Fig.~\ref{fig:charge}(b)]. Hence, the charging speed is governed by the dissipative dynamics rather than by the statistics of the charger.

\subsection{Charging with the charger in a coherent state}

\begin{figure*}[t]
\includegraphics[width=\textwidth]{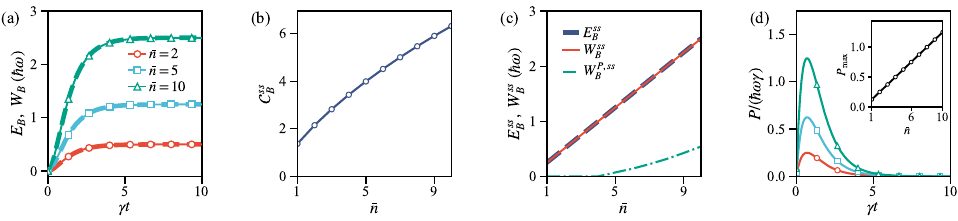}
\caption{Charging performance for a coherent-state charger with mean photon number
$\bar n$. (a)~Stored energy $E_B(t)$ (thick dashed) and ergotropy $W_B(t)$ (thin solid, with
markers) versus rescaled time $\gamma t$, for $\bar n=2$, $5$, and $10$. (b)~Steady-state
coherence $\mathcal{C}_B^{ss}$ versus $\bar n$. (c)~Steady-state stored energy $E_B^{ss}$ (thick
dashed), ergotropy $W_B^{ss}$ (thin solid), and its incoherent part $W_B^{P,ss}$ (dash-dotted)
versus $\bar n$. (d)~Charging power $P(t)=dE_B/dt$ for the same three $\bar n$, with the colors
and markers of (a); the inset shows the peak power $P_{\max}$ versus $\bar n$. Energies are in units of $\hbar\omega$ and
powers in units of $\hbar\omega\gamma$.}
\label{fig:coh}
\end{figure*}

For a Fock-state charger, the steady-state battery is diagonal in the Fock basis; no coherence
survives, and the ergotropy is carried entirely by population inversion. Introducing coherence
should therefore enhance the QB performance. We thus consider a battery prepared in a coherent
state with mean photon number $\bar n=\vert\alpha\vert^{2}$. In the canonical-mode basis, the
initial state is
\begin{equation}
\vert\psi(0)\rangle=\vert\alpha\rangle_{c}\vert0\rangle_{b}
=\left\vert\frac{\alpha}{\sqrt2}\right\rangle_{A}
\left\vert-\frac{\alpha}{\sqrt2}\right\rangle_{B},
\end{equation}
the dark state mode $B$ remains steady, and the decay mode $A$ undergoes single-photon loss. The
evolution state is
\begin{equation}
\rho(t)=\left\vert\frac{\alpha e^{-\gamma t}}{\sqrt2}\right\rangle_{A}
\left\langle\frac{\alpha e^{-\gamma t}}{\sqrt2}\right\vert\otimes
\left\vert-\frac{\alpha}{\sqrt2}\right\rangle_{B}
\left\langle-\frac{\alpha}{\sqrt2}\right\vert,
\end{equation}
leading to the corresponding steady state
\begin{equation}
\rho_B^{ss}=\vert-\alpha/2\rangle_{b}\langle-\alpha/2\vert.
\end{equation}
The stored energy is
\begin{equation}
E_B^{ss}=\vert\alpha\vert^{2}\hbar\omega/4=\bar n\hbar\omega/4,
\end{equation}
which scales linearly with the initial photon number. For a coherent state, the passive state is the
vacuum state, so the ergotropy naturally saturates the stored energy (see Appendix~\ref{app:coherent} for the
derivation). Numerical results are shown in
Fig.~\ref{fig:coh}.

Fig.~\ref{fig:coh}(a) shows that the stored energy and the ergotropy grow and saturate at their
steady-state values, similar to the Fock case. Here, however, the ergotropy saturates at the stored
energy. The ergotropy is dominated by the steady-state coherence rather than by population
inversion. Fig.~\ref{fig:coh}(b) confirms this: the steady-state coherence remains nonzero and grows
with $\bar n$. The corresponding steady-state energies and ergotropy are shown in
Fig.~\ref{fig:coh}(c), where the incoherent part (population inversion) of the ergotropy is small
and vanishes for $\bar n\le4$. These results further show that, in contrast to the Fock case, the
dominant contribution to the ergotropy is the coherence inherited from the charger's initial state.

Finally, we compute the time-dependent power and extract its maximum for different $\bar n$. The
stored energy as a function of the evolution time is
\begin{equation}
E_B(t)=\frac{\bar n}{4}\hbar\omega\left(1-e^{-\gamma t}\right)^{2},
\end{equation}
and the charging power follows from its derivative,
\begin{equation}
P(t)=\frac{\bar n\hbar\omega\gamma}{2}e^{-\gamma t}\left(1-e^{-\gamma t}\right),
\end{equation}
which peaks at $\gamma t=\ln2$ with $P_{\max}=\bar n\hbar\omega\gamma/8$. The results are again
similar to the Fock case, with $P_{\max}\propto\bar n$ [inset of Fig.~\ref{fig:coh}(d)]. Thus, the
peak power is governed by the charging dynamics, with the same linear scaling in $\bar n$ as the
Fock-state charger.

\section{Superextensive charging by two-photon dissipation}

Neither a Fock-state nor a coherent-state charger exhibits superextensive charging: the peak power grows only linearly with the mean photon number, so that the charging is merely extensive. We therefore consider
\begin{equation}\label{master2}
\dot\rho(t)=-i\omega\left[A^{\dagger}A+B^{\dagger}B,\rho(t)\right]+\gamma_2\mathcal{D}[A^2]\rho(t),
\end{equation}
in which two-photon loss acts on the canonical mode $A$. This choice preserves dark-state protection, because $A^{2}$ annihilates the dark states. In contrast, a symmetric channel $\mathcal{D}[b^{2}+c^{2}]=\mathcal{D}[A^{2}+B^{2}]$ would damp the dark mode as well and destroy the protection. 

\begin{figure*}[t]
\includegraphics[width=\textwidth]{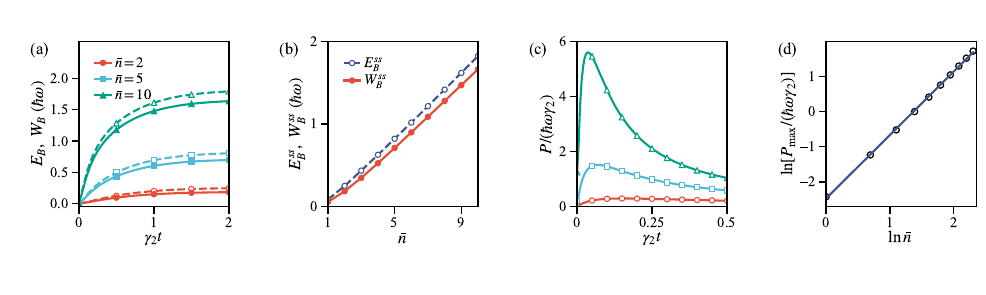}
\caption{Charging performance for a coherent-state charger under two-photon dissipation
$\gamma_2\mathcal{D}[A^2]$. (a)~Stored energy $E_B(t)$ (dashed) and ergotropy $W_B(t)$ (solid,
with markers) versus rescaled time $\gamma_2t$, for $\bar n=2$, $5$, and $10$. (b)~Steady-state
stored energy $E_B^{ss}$ (dashed) and ergotropy $W_B^{ss}$ (solid) versus $\bar n$.
(c)~Charging power $P(t)=dE_B/dt$ for the same three $\bar n$. (d)~Peak power $P_{\max}$ versus
$\bar n$ on a logarithmic scale; symbols are the numerical data and the line is a linear fit,
$P_{\max}\propto\bar n^{1.81}$. Energies are in units of $\hbar\omega$ and powers in units of
$\hbar\omega\gamma_2$.}
\label{fig:2ph}
\end{figure*}

For a coherent-state charger, the situation is only partly analogous: $B$ remains decoupled and only $A$ is damped, but $A$ no longer evolves into a coherent state, because $A^{2}$ does not map coherent states onto coherent states. Consequently, the passive state of the battery is no longer the vacuum, the ergotropy no longer saturates the stored energy, and the coherence remaining in $A$ renders the charging nonmonotonic in time. We therefore evaluate the performance numerically, as shown in Fig.~\ref{fig:2ph}.

Figure~\ref{fig:2ph}(a) shows that the stored energy and the ergotropy saturate
within $\gamma_2t\approx2$, with the saturation time decreasing as $\bar n$ increases. Both
steady-state values grow approximately linearly with $\bar n$ [Fig.~\ref{fig:2ph}(b)], reaching
$E_B^{ss}=1.82\,\hbar\omega$ and $W_B^{ss}=1.66\,\hbar\omega$ at $\bar n=10$. The ergotropy
therefore amounts to $74\%$ of the stored energy at $\bar n=2$ and to $91\%$ at $\bar n=10$.
Compared with the single-photon channel, the same initial state stores slightly less energy
($0.18\,\bar n\hbar\omega$ instead of $0.25\,\bar n\hbar\omega$) but delivers it much faster. The
charging power [Fig.~\ref{fig:2ph}(c)] peaks at $\gamma_2t\simeq0.4/\bar n$ and grows faster than
linearly with $\bar n$. Figure~\ref{fig:2ph}(d) displays the peak power on a logarithmic scale: a
linear fit gives $P_{\max}\propto\bar n^{1.81}$, and the local slope rises from $1.7$ to $1.9$
across the plotted range, approaching the superextensive exponent $2$.

\section{Experimental realization}
\begin{figure}[t]
\includegraphics[width=\columnwidth]{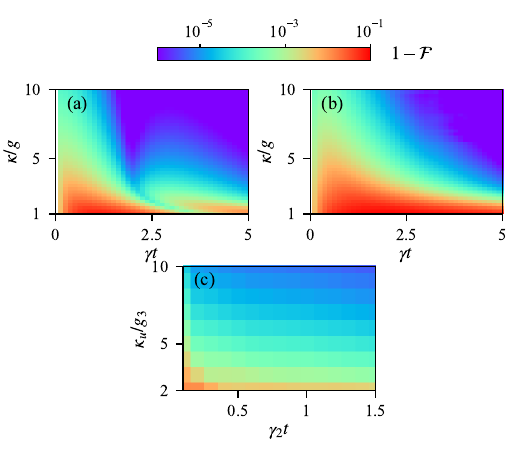}
\caption{Deviation $1-\mathcal{F}$ between the full three-mode dynamics and the effective
single-mode description, shown on a logarithmic color scale versus the loss-to-coupling ratio and
the evolution time. (a)~Fock-state charger with $n=3$ and (b)~coherent-state charger with
$\bar n=3$, for the single-photon channel, where the axes are $\kappa/g$ and $\gamma t$.
(c)~Coherent-state charger for the two-photon channel of Eq.~(\ref{eq:effme2}), where the axes are
$\kappa_{u}/g_{3}$ and $\gamma_{2}t$. The agreement improves rapidly with the loss-to-coupling ratio
and is exact at $\gamma t=0$, where both descriptions start from the same state.}
\label{fig:fidmap}
\end{figure}

Simulating a reservoir that couples simultaneously to two bosonic modes is challenging experimentally. We instead construct such a reservoir explicitly \cite{arXiv.2611.06712}: two
high-$Q$ resonators coupled to a common low-$Q$ resonator. In the frame rotating at resonators' resonant frequency $\omega$, the
Hamiltonian reads
\begin{equation}
      H_{r}=g\left[(b^{\dagger}+c^{\dagger})d+(b+c)d^{\dagger}\right],
\end{equation}
where $b$ and $c$ ($b^{\dagger}$ and $c^{\dagger}$) are the annihilation (creation) operators of the two
high-$Q$ resonators, respectively, $d$ ($d^{\dagger}$) is that of the low-$Q$ resonator, and $g$ is the coupling
strength. In terms of the canonical mode $A=(b+c)/\sqrt{2}$, the Hamiltonian becomes
\begin{equation}
      H_{r}'=\sqrt{2}g\left(A^{\dagger}d+Ad^{\dagger}\right).
\end{equation}
Neglecting the small losses of the high-$Q$ resonators, the master equation is
\begin{equation}
      \frac{\partial\rho}{\partial t}=-i\left[H_{r}',\rho\right]+\kappa\mathcal{D}[d]\rho,
\end{equation}
where $\kappa$ is the loss rate of the low-$Q$ resonator. For $\kappa\gg g$, the low-$Q$ resonator is adiabatically eliminated via a Purcell-type
decay, and the dynamics reduces to that of a single lossy mode $A$,
\begin{equation}
      \frac{\partial\rho}{\partial t}=2\gamma_{\rm eff}\mathcal{D}[A]\rho,
\label{eq:effme}
\end{equation}
with $\gamma_{\rm eff}=g^{2}/\kappa$. To verify the accuracy of this elimination, we simulate the full
dynamics and quantify the agreement with the fidelity
\begin{equation}
      \mathcal{F}=\left({\rm tr}\sqrt{\sqrt{\rho_{i}}\rho_{e}\sqrt{\rho_{i}}}\right)^{2},
\end{equation}
where $\rho_i$ and $\rho_e$ are the density matrices of the ideal (adiabatically eliminated) model and full model, respectively. Figure~\ref{fig:fidmap} shows $1-\mathcal{F}$ on a logarithmic color scale as functions of $\kappa/g\in[1,10]$ and the evolution time $\gamma t$, for a charger initialized in a Fock state (a) and a coherent state (b) with initial (mean) photon number $3$. The two descriptions agree well over the entire range and
become indistinguishable as $\kappa/g$ increases: the deviation decreases monotonically from about $10^{-1}$ at
$\kappa/g=1$ to $10^{-3}$ near $\kappa/g\simeq5$ and to a few $10^{-4}$ at $\kappa/g=10$; it vanishes at
$\gamma t=0$, where both models start from the same state. The coherent case is slightly more demanding
than the Fock case at the same $\kappa/g$, because its broader photon-number distribution samples the
correction terms more strongly. These results confirm that Eq.~(\ref{eq:effme}) reproduces the full
dynamics once $\kappa\gtrsim5g$.

The same architecture extends naturally to the two-photon channel of
Eq.~(\ref{master2}). The linear coupling is replaced by a three-wave-mixing element: the two
high-$Q$ resonators are coupled symmetrically to a common superconducting nonlinear asymmetric inductive element (SNAIL), whose flux-tunable cubic
nonlinearity supplies the $\varphi^{3}$ term absent in a symmetric Josephson junction \cite{APL.110.222603,PhysRevApplied.11.054060}; the derivation is given in Appendix~\ref{app:3wm}. With a lossy buffer mode $u$ and a pump at
$\omega_{p}=2\omega_{A}-\omega_{u}$, the rotating-frame interaction picture reads
\begin{equation}
H_{3\rm wm}=g_{3}\left(A^{2}u^{\dagger}+A^{\dagger2}u\right),
\label{eq:3wm}
\end{equation}
where $g_{3}$ is the coupling strength between mode $A$ and buffer mode $u$, and the cubic coefficient is modulated by the pump. For $\kappa_{u}\gg g_{3}$, eliminating the buffer as in Eq.~(\ref{eq:effme}), yields
\begin{equation}
\frac{\partial\rho}{\partial t}=\gamma_{2}\mathcal{D}[A^{2}]\rho,\qquad
\gamma_{2}=\frac{g_{3}^{2}}{\kappa_{u}},
\label{eq:effme2}
\end{equation}
the two-photon counterpart of the single-photon result. This is the mechanism used to
stabilize cat qubits, where two-photon loss is engineered with the same buffer-and-pump
construction \cite{NatPhys.16.509,PhysRevX.15.011070}. We verify this elimination numerically in Fig.~\ref{fig:fidmap}(c): for a coherent-state charger, the deviation $1-\mathcal{F}$ between the full three-mode dynamics and
Eq.~(\ref{eq:effme2}) decreases from $1.2\times10^{-2}$ at $\kappa_{u}/g_{3}=2$ to
$1.0\times10^{-3}$ at $\kappa_{u}/g_{3}=5$ and to $3\times10^{-5}$ at $\kappa_{u}/g_{3}=10$.

\section{Discussion and outlook}
We have studied the charging of an open bosonic QB whose battery mode $b$ and charger mode $c$ interact only through a common dissipative reservoir. In the
canonical-mode basis, the symmetric combination $A=(b+c)/\sqrt{2}$ decays at rate $2\gamma$, whereas
the antisymmetric combination $B=(b-c)/\sqrt{2}$ is decoupled from the reservoir; the
dark-mode occupation is therefore conserved, and the battery is protected against dissipation. For a Fock-state charger, we obtained the exact steady state of the battery, which yields a stored
energy growing linearly with $n$. Because this state is diagonal in the Fock basis, no coherence
survives, and the ergotropy stems solely from the population inversion. To further improve the QB
performance, we introduced coherence by preparing the charger in a coherent
state with mean photon number $\bar n$. The extractable work then saturates at the
stored energy, with the incoherent contribution being small and vanishing for $\bar n\le4$. In both cases, the
charging power peaks at $\gamma t\simeq0.7$ and scales linearly with the photon number,
$P_{\max}\propto n$ for the Fock charger and $P_{\max}\propto\bar n$ for the coherent one. For the two-photon channel, the stored energy still grows linearly with $\bar n$, while the charging time decreases as $1/\bar n$, so that the peak power becomes superextensive and approaches $\bar n^{2}$. 

Several directions remain. Squeezing is a natural resource to explore: a squeezed state
carries photon-number correlations that a coherent state lacks and may further enhance the
extractable work and the charging rate. A finite reservoir temperature, together with implementations in circuit QED or photonic platforms with engineered two-mode dissipation, would connect the idealized model
studied here to realistic devices. More broadly, our results show that the dark subspace of a
multimode bosonic system can be exploited as a passive resource for protecting quantum energy
storage, and that the performance of such a battery can be pushed by adding excitations rather than
particles. We expect that this combination of few-body implementation and dissipation-immune
operation will stimulate further work on bosonic open QBs and, more generally, on quantum
thermodynamics with engineered bosonic environments.

\section*{Data Availability Statement}
The data that support this study are available from the corresponding author upon reasonable request.

\section*{Acknowledgments}
This work was supported by the National Natural Science Foundation of China (Grants No. 12474356, No. 12475015, No. 12274080, No. 12505016), Quantum Science and Technology-National Science and Technology Major Project (Grant No. 2021ZD0300200), and the Natural Science Foundation of Fujian Province (Grants No. 2025J01465).

\appendix
\section{Steady state of the battery for a Fock-state charger}\label{app:steady}

For a Fock-state charger, the steady state can be derived analytically. We expand the initial state in the Fock basis of the canonical modes,
\begin{equation}
\vert\psi(t=0)\rangle=\vert0\rangle_{b}\vert n\rangle_{c}=\sum_{k=0}^{n}C_{k}\vert n-k,k\rangle_{A,B},
\end{equation}
with $C_{k}=\langle n-k,k\vert\psi(t=0)\rangle$. The canonical-mode basis states are
\begin{equation}\label{eq:n-k}
      \begin{split}
            \vert n-k,k\rangle_{A,B}
            ={}&\frac{(A^{\dagger})^{n-k}}{\sqrt{(n-k)!}}\frac{(B^{\dagger})^{k}}{\sqrt{k!}}\vert vac\rangle\\
            ={}&\frac{1}{\sqrt{2^{n}}}
            \frac{\sqrt{(j+i)!(n-i-j)!k!(n-k)!}}{i!j!(k-j)!(n-k-i)!}\\
            &\times\sum_{i=0}^{n-k}\sum_{j=0}^{k}(-1)^{k-j}
            \vert n-i-j\rangle_{c}\vert i+j\rangle_{b},
      \end{split}
\end{equation}
where $A^{\dagger}=(b^{\dagger}+c^{\dagger})/\sqrt{2}$ and
$B^{\dagger}=(b^{\dagger}-c^{\dagger})/\sqrt{2}$. Introducing $m=i+j$, and hence $i=m-j$,
we rewrite Eq.~(\ref{eq:n-k}) as
\begin{equation}
      \begin{split}
            \vert n-k,k\rangle_{A,B}
            ={}&\sum_{m=0}^{n}\sqrt{\frac{m!(n-m)!}{2^{n}\,k!\,(n-k)!}}\\
            &\times\sum_{j}(-1)^{k-j}\binom{k}{j}\binom{n-k}{m-j}\\
            &\times\vert n-m\rangle_{c}\vert m\rangle_{b}\\
            ={}&\sum_{m=0}^{n}\sqrt{\frac{1}{2^{n}}\binom{n}{k}\binom{n}{m}}\,(-1)^{k}K_{m}(k)\\
            &\times\vert n-m\rangle_{c}\vert m\rangle_{b},
      \end{split}
\end{equation}
where $K_{m}(k)$ is the Krawtchouk polynomial,
\begin{equation}
      K_{m}(k)=\frac{1}{\binom{n}{m}}\sum_{j=0}^{n}(-1)^{j}\binom{k}{j}\binom{n-k}{m-j},
\end{equation}
which satisfies the orthogonality relation
\begin{equation}
      \begin{split}
      &\sum_{m=0}^{n}\binom{n}{m}K_{m}(k')K_{m}(k)=\frac{2^{n}}{\binom{n}{k}}\,\delta_{kk'},\\
      &\sum_{k=0}^{n}\binom{n}{k}K_{m}(k)=2^{n}\delta_{m0}.
      \end{split}
\end{equation}
Using these relations, we obtain
\begin{equation}
      C_{k}=\sqrt{\frac{1}{2^{n}}\binom{n}{k}}\,(-1)^{k}K_{0}(k)=\sqrt{\frac{1}{2^{n}}\binom{n}{k}}\,(-1)^{k},
\end{equation}
and hence
\begin{equation}
      \rho_{ss}=\frac{1}{2^{n}}\sum_{k}\binom{n}{k}\vert0,k\rangle_{A,B}\langle0,k\vert .
\end{equation}
Tracing out the charger leaves
\begin{equation}
      \mathrm{Tr}_{c}\left[\vert0,k\rangle_{A,B}\langle0,k\vert\right]
      =\frac{1}{2^{k}}\sum_{j=0}^{k}\binom{k}{j}\vert j\rangle_{b}\langle j\vert,
\end{equation}
so that the steady state of the battery is
\begin{equation}
      \rho_{B}=\sum_{j=0}^{n}\binom{n}{j}\sum_{k=j}^{n}\frac{1}{2^{n+k}}\binom{n-j}{k-j}\vert j\rangle_{b}\langle j\vert .
\end{equation}
Setting $m=k-j$, we find
\begin{equation}
      \begin{split}
            \rho_{B}
            ={}&\sum_{j=0}^{n}\binom{n}{j}\frac{1}{2^{n+j}}
            \left[\sum_{m=0}^{n-j}\binom{n-j}{m}\left(\frac{1}{2}\right)^{m}\right]\vert j\rangle_{b}\langle j\vert\\
            ={}&\sum_{j=0}^{n}\binom{n}{j}\frac{1}{2^{n+j}}\left(\frac{3}{2}\right)^{n-j}\vert j\rangle_{b}\langle j\vert\\
            ={}&\frac{1}{4^{n}}\sum_{j=0}^{n}\binom{n}{j}3^{\,n-j}\vert j\rangle_{b}\langle j\vert
            =\sum_{j=0}^{n}p_{j}\vert j\rangle_{b}\langle j\vert .
      \end{split}
\end{equation}
Introducing the generating function $G(t)=\sum_{j}p_{j}t^{j}$, the stored energy follows as
\begin{equation}
      E^{ss}_{B}=\hbar\omega\langle b^{\dagger}b\rangle=\hbar\omega\sum_{j}j\,p_{j}
      =\hbar\omega\left.\frac{\partial G(t)}{\partial t}\right\vert_{t=1}=\frac{n}{4}\hbar\omega .
\end{equation}
Thus the stored energy of the QB is proportional to the photon number of the charger.
\section{Steady state of the battery for a coherent-state charger}\label{app:coherent}

For a coherent-state charger, the charging dynamics is exactly solvable: the initial state factorizes
in the canonical-mode basis, and single-mode damping maps coherent states onto coherent
states. Using Eq.~(\ref{canonical}) and $[A,B]=0$, the initial state becomes
\begin{equation}
\vert\alpha\rangle_{c}\vert0\rangle_{b}
=\left\vert\frac{\alpha}{\sqrt{2}}\right\rangle_{A}
\left\vert-\frac{\alpha}{\sqrt{2}}\right\rangle_{B},
\end{equation}
so that both canonical modes start in coherent states of equal amplitude. The relative sign follows
from Eq.~(\ref{canonical}) and does not affect any quantity evaluated below.

A mode damped at rate $\gamma$ obeys
$\dot\rho=\gamma\left(a\rho a^{\dagger}-\frac12\{a^{\dagger}a,\rho\}\right)$. For
$\rho=\vert\beta\rangle\langle\beta\vert$, the relations $a\vert\beta\rangle=\beta\vert\beta\rangle$
and $\langle\beta\vert a^{\dagger}=\beta^{*}\langle\beta\vert$ reduce the right-hand side to
$\gamma\left[\vert\beta\vert^{2}\rho-\frac12\beta a^{\dagger}\rho-\frac12\beta^{*}\rho a^{\dagger}\right]$,
which is reproduced identically by $\beta(t)=\beta e^{-\gamma t}$ on account of
$\partial_{\beta}\vert\beta\rangle=\left(a^{\dagger}-\beta^{*}/2\right)\vert\beta\rangle$, i.e.
\begin{equation}
\vert\beta\rangle\langle\beta\vert\longmapsto
\vert\beta e^{-\gamma t}\rangle\langle\beta e^{-\gamma t}\vert ,
\end{equation}
so that state remains pure. Since mode $B$ is decoupled, the two-mode state at time $t$ is
\begin{equation}
\begin{split}
\rho(t)={}&\left\vert\frac{\alpha}{\sqrt{2}}e^{-\gamma t}\right\rangle_{A}
\left\langle\frac{\alpha}{\sqrt{2}}e^{-\gamma t}\right\vert\\
&\otimes\left\vert-\frac{\alpha}{\sqrt{2}}\right\rangle_{B}
\left\langle-\frac{\alpha}{\sqrt{2}}\right\vert ,
\end{split}
\end{equation}
Transforming back with Eq.~(\ref{canonical}) gives $\rho(t)=\vert\gamma_b\rangle\langle\gamma_b\vert
\otimes\vert\gamma_c\rangle\langle\gamma_c\vert$ with
\begin{equation}
\gamma_b(t)=\frac{\alpha}{2}\left(e^{-\gamma t}-1\right),\qquad
\gamma_c(t)=\frac{\alpha}{2}\left(e^{-\gamma t}+1\right).
\end{equation}
Since $\gamma_b\to-\alpha/2$ and $\gamma_c\to\alpha/2$ at long times, the steady-state density
matrix of the battery is
\begin{equation}
\rho^{B}_{ss}=\vert-\alpha/2\rangle\langle-\alpha/2\vert ,
\end{equation}
a pure coherent state, in contrast to the diagonal mixture obtained for a Fock-state charger.

All figures of merit follow analytically. With $\bar n=\vert\alpha\vert^{2}$,
\begin{equation}
\langle n_b(t)\rangle=\frac{\bar n}{4}\left(1-e^{-\gamma t}\right)^{2},\quad
E_B(t)=\frac{\bar n}{4}\hbar\omega\left(1-e^{-\gamma t}\right)^{2},
\end{equation}
\begin{equation}
P(t)=\frac{\bar n\hbar\omega\gamma}{2}e^{-\gamma t}\left(1-e^{-\gamma t}\right),
\end{equation}
which peaks at $\gamma t=\ln2$ with $P_{\max}=\bar n\hbar\omega\gamma/8$, consistent with
Fig.~\ref{fig:coh}(d). Because $\rho_B(t)$ is pure, its passive state is the vacuum and
$W_B(t)=E_B(t)$ at all times: the entire stored energy is extractable. The coherence is that of a
coherent state with a Poissonian photon distribution of mean $\mu(t)=\vert\gamma_b(t)\vert^{2}$,
\begin{equation}
\mathcal{C}_B(t)=\left(\sum_{n}\sqrt{p_n(t)}\right)^{2}-1,\qquad
p_n(t)=e^{-\mu}\frac{\mu^{n}}{n!},
\end{equation}
and $\mathcal{C}^{ss}_B\simeq\sqrt{\bar n}$ for $\bar n\ll4$.

The incoherent part of the ergotropy vanishes whenever the diagonal of $\rho_B$ is already passive,
i.e.\ when the Poisson distribution decreases monotonically with $n$, which requires $\mu\le1$.
Since $\mu\to\bar n/4$ at long times, $W^{P,ss}_B=0$ for $\bar n\le4$ and becomes finite above this
threshold, as observed in Fig.~\ref{fig:coh}(c). By contrast, the Fock-state charger yields the
mixed state of Eq.~(\ref{fockdist}), whose ergotropy originates solely from the population inversion.

\section{Engineering of the two-photon channel}\label{app:3wm}

The two-photon channel of Eq.~(\ref{master2}) requires a nonlinear element with an odd-order
potential. The two resonators are coupled symmetrically to a SNAIL \cite{APL.110.222603}, a loop of
$n$ large junctions of energy $E_{J}$ and one smaller junction of energy $\alpha E_{J}$, threaded
by an external flux $\Phi_{\rm ext}$, whose inductive energy is
\begin{equation}
U(\varphi)=-\alpha E_{J}\cos\varphi
-nE_{J}\cos\left(\frac{2\pi\Phi_{\rm ext}/\Phi_{0}-\varphi}{n}\right).
\end{equation}
Expanding about the flux-dependent minimum,
$U_{\rm eff}/E_{J}=c_{2}\tilde\varphi^{2}+c_{3}\tilde\varphi^{3}
+c_{4}\tilde\varphi^{4}+\cdots$, yields a cubic term only for $n\ge2$: a SQUID ($n=1$) reduces to a
pure cosine about its minimum and has $c_{3}=0$ \cite{APL.110.222603}. The device is biased at the
working point where the quartic term is cancelled while the cubic term survives, $c_{3}\ne0$ and
$c_{4}=0$ \cite{APL.110.222603,PhysRevApplied.11.054060}. The pump is applied as a modulation of
the flux threading the loop, 
\begin{equation}
\Phi_{\rm ext}(t)=\bar\Phi_{\rm ext}+\delta\Phi\cos\omega_{p}t,
\end{equation} 
so that
\begin{equation}
c_{3}(t)=c_{3}+c_{3}'\delta\Phi\cos\omega_{p}t, \ c_{3}'=\partial c_{3}/\partial\Phi_{\rm ext}.
\end{equation}
A lossy buffer mode $u$ completes the circuit, and
the phase across the element is
\begin{equation}
\varphi=\sqrt{2}\varphi_{A}\left(A+A^{\dagger}\right)
+\varphi_{u}\left(u+u^{\dagger}\right),
\end{equation}
with the antisymmetric combination $B$ absent by symmetry. Keeping, in the expansion of
Eq.~(C1), the terms that rotate at the pump frequency, with the resonance condition $\omega_{p}=2\omega_{A}-\omega_{u}$, one obtains
\begin{equation}
H_{3\rm wm}=g_{3}\left(A^{2}u^{\dagger}+A^{\dagger2}u\right),
\qquad g_{3}\simeq\tfrac{1}{2}E_{J}c_{3}'\delta\Phi\,\varphi_{A}^{2}\varphi_{u},
\end{equation}
where the collective operator $A^{2}$ appears because both resonators enter the phase with the same
coefficient. A symmetric junction ($c_{3}=0$) would yield only a quartic term and no
three-wave mixing, hence the need for the SNAIL.

Eliminating the buffer in the limit $\kappa_{u}\gg g_{3}$ then yields the effective
dynamics used in the main text, Eq.~(\ref{master2}), with
\begin{equation}
\frac{\partial\rho}{\partial t}=\gamma_{2}\mathcal{D}[A^{2}]\rho,\qquad
\gamma_{2}=\frac{g_{3}^{2}\kappa_{u}}{\kappa_{u}^{2}+\Delta_{u}^{2}},
\end{equation}
where $\Delta_{u}\equiv \omega_p-2\omega_A+\omega_u$ is the residual detuning of the buffer in the rotating frame. At resonance, $\Delta_{u}=0$ and $\gamma_{2}=g_{3}^{2}/\kappa_{u}$, the two-photon analog of $\gamma_{\rm eff}=g^{2}/\kappa$.

\makeatletter
\immediate\write\@auxout{\string\citation{TitleControl}}
\makeatother
\bibliographystyle{apsrev4-2}
\bibliography{refs}

\end{document}